\documentclass[12pt]{article}

\usepackage{amssymb}

\begin{document}

\begin{titlepage}

\begin{flushright}
arXiv:1802.05838
\end{flushright}
\vskip 2.5cm

\begin{center}
{\Large \bf Mode Analysis for Energetics of a Moving Charge\\
In Lorentz- and CPT-Violating Electrodynamics}
\end{center}

\vspace{1ex}

\begin{center}
{\large Richard DeCosta and Brett Altschul\footnote{{\tt baltschu@physics.sc.edu}}}

\vspace{5mm}
{\sl Department of Physics and Astronomy} \\
{\sl University of South Carolina} \\
{\sl Columbia, SC 29208} \\
\end{center}

\vspace{2.5ex}

\medskip

\centerline {\bf Abstract}

\bigskip

In isotropic but Lorentz- and CPT-violating electrodynamics, it is known that a charge in uniform
motion does not lose any energy to Cerenkov radiation. This presents a puzzle, since the radiation
appears to be kinematically allowed for many modes. Studying the Fourier transforms of the most
important terms in the modified magnetic field and Poynting vector, we confirm the vanishing
of the the radiation rate. Moreover, we show that the Fourier transform of the field
changes sign between small and large wave numbers. This enables modes with very long wavelengths
to carry negative energies, which cancel out the positive energies carried away by modes
with shorter wavelengths. This cancelation had previously been inferred but never explicitly
demonstrated.

\bigskip

\end{titlepage}

\newpage

\section{Introduction}

Symmetry has proven to be a key topic in our understanding of modern physics.
For instance, many transformations that initially appeared to be symmetries of the
standard model of particle physics,
but which ultimately proved not to be exact symmetry operations, have provided important
insights into the structure of the theory. 
Whatever new physics exists beyond the standard model might involve additional
interesting forms of symmetry breaking.
Among the most extreme symmetry violations that might be possible are
violations of Lorentz and CPT invariances. These symmetries are related to
very basic properties of the theory, describing isotropy, boost invariance, and
Hermiticity of the Hamiltonian. Both types of symmetry are also building
blocks of both the standard model and the general theory of relativity, yet in the
ultimate quantum gravity theory, these symmetries might not hold precisely.
Indeed, many theories that have been proposed in attempts to described the schematic
properties of quantum gravity seem to have regimes in which Lorentz and CPT invariances
may not hold.

Experimental searches for violations of fundamental symmetries
can provide important information about the character of new physics. However,
even if there is no Lorentz or CPT violation in nature, studying exotic
field theories can help us to understand the general character of quantum theory.
Such theories may provide fundamental new insights about the kinds of behaviors that
are permitted in the general field theory framework.

A natural formalism for addressing these kinds of questions is effective field theory.
The general effective field theory that delineates Lorentz- and CPT-violating additions
to the standard model is known as the standard model extension (SME), and it
has been the focus of extensive phenomenalistic study for the past two decades.
The action for the SME can be constructed from all operators built up out of the usual
standard model fields~\cite{ref-kost1,ref-kost2}. In the standard model, these operators
are subject to the requirement that they be Lorentz scalars, but in the SME
that requirement is absent. As result, the number of possible operators is much larger
than in the Lorentz-invariant theory.
For most practical calculations, the minimal SME is used; it contains only 
those operators that are local, power counting
renormalizable, and gauge invariant.
Most experimental test of Lorentz and CPT symmetries are now used to place
constraints on minimal SME parameters.

Not all forms of Lorentz violation are equally exotic. While all reasonable CPT-violating
theories are also Lorentz-violating~\cite{ref-greenberg}, the reverse is not true. There are
Lorentz-violating operators that are even under CPT. Moreover, there can
be even more subtle connections between Lorentz symmetry and properties like gauge
invariance. The electromagnetic Cherm-Simons term in the minimal SME Lagrange density
is not gauge invariant; it depends explicitly on the potentials, not just on
field strengths. However the integrated action is nonetheless gauge invariant,
and the equations of motions only involve $\vec{E}$ and $\vec{B}$, not
$\vec{A}$ and $A_{0}$. The subtleties associated with the implementation of this
kind of term in a quantum theory had provoked quite a bit of controversy in the
past; there was significant debate about the right way of calculating radiative
corrections to a bare Chern-Simons term~\cite{ref-coleman,
ref-jackiw1,ref-victoria1,ref-chung1,ref-andrianov,ref-altschul1}.
Moreover, the Chern-Simons theory also suffers from long-wavelength instabilities.

The Chern-Simons term is one of the most interesting terms in the SME, from a
theoretical point of view. However, it is also one of the easiest terms to
bound in practice. The term affects the propagation of the left and
right circular polarization modes of the electromagnetic field differently. The
differences between the modes' dispersion relations lead to vacuum birefringence.
The distinctive birefringence signature has not been seen,
even for waves originating at cosmological distances~\cite{ref-carroll1,ref-kost21,ref-mewes5}.
The lack of birefringence has been used to place exceedingly tight bounds on
the size of the real-world Chern-Simons term.

Several previous analyses have looked at another peculiar feature
of the Chern-Simons theory---the possibility of vacuum Cerenkov emission. The Cerenkov process
is normally forbidden in vacuum by energy-momentum conservation and Lorentz invariance.
However, if particles can possess Lorentz-violating energy-momentum relations, it may be possible for
charged particles to move faster than the the phase speed of light. Since the Chern-Simons
term changes the dispersion relations for electromagnetic waves, including slowing one
polarization down, vacuum Cerenkov radiation is a natural possibility in this theory.
However, there is an iterative algorithm for
determining the electric and magnetic fields of a moving point charge in the modified theory, and 
studies of the symmetry properties of this algorithm have showed that in the case of a
timelike Chern-Simons coefficient, there is zero radiation power loss from a uniformly moving
charge~\cite{ref-schober1}.

This vanishing of the total Cerenkov radiation rate leaves a number of puzzles associated with it.
This paper will clarify the structure of the relevant field components in Fourier space, making
explicit a cancelation that has previously only been indirectly inferred. The paper is organized as follows:
Section~\ref{sec-theory} introduces the action of the Chern-Simons theory and the structure of the
modified energy-momentum tensor. In section~\ref{sec-fourier}, we calculate the Fourier transforms of the
magnetic field and the Poynting vector at the lowest relevant orders. Section~\ref{sec-cancel} shows
how there can be a cancelation between short-wavelength modes carrying positive energies and long-wavelength
modes that actually carry negative total energies. Finally, section~\ref{sec-concl} summarizes our conclusions
about the interpretation of the paper's results.

\section{Lorentz-Violating Chern-Simons Theory}
\label{sec-theory}

The theory we will consider is rotation invariant (in a preferred frame),
but the CPT and Lorentz boost symmetries are broken. There are also
CPT-even and anisotropic forms of Lorentz violation in the photon
sector of the SME. In fact, the minimal SME electromagnetic Lagrange density, including all
terms that can be constructed solely out of photon
operators, is
\begin{equation}
{\cal L}=-\frac{1}{4}F^{\mu\nu}F_{\mu\nu}
 -\frac{1}{4}k_{F}^{\mu\nu\rho\sigma}F_{\mu\nu}F_{\rho\sigma}
 +\frac{1}{2}k_{AF}^{\mu}\epsilon_{\mu\nu\rho\sigma}F^{\nu\rho}A^{\sigma}
-j^{\mu}A_{\mu}.
\end{equation}
The CPT-even terms are those parameterized by the nineteen independent $k_{F}$
coefficients. They have many interesting possible effects, but we shall neglect
them here. The four $k_{AF}$ coefficients multiply the possible CPT-odd operators,
with the time component of $k_{AF}$ multiplying the only one that is also isotropic.
The structure of the $k_{AF}$ term is a four-dimensional, Lorentz-violating
generalization of a Chern-Simons term.

So we shall consider a strictly timelike vector
$k_{AF}^{\mu}=(k,\vec{0}\,)$. This makes the $k_{AF}$ term proportional to just
$\vec{A}\cdot\vec{B}$. While this Lorentz-violating term is fairly simple in form,
it appears to have many problematic properties. One of the most obvious
difficulties is that the dispersion relation for circularly polarized photons of momentum
$\vec{Q}$ becomes
$\omega_{\pm}^{2}=Q(Q\mp 2k)$; the sign of the unusual term is the negative of the
helicity of the mode. It is clear that for waves with very long wavelengths $Q<|2k|$, one set of the
helicity modes will have imaginary frequencies. This naturally can give rise to runaway solutions,
which grow exponentially in time. It is possible to avoid these runaway modes by selecting
the Green's functions for the theory in a very special way; \cite{ref-carroll1} exhibits a
Green's function that has only real frequency components, but at the cost of using
acausal boundary conditions. As a result, a charged particle will
start to emit radiation before it actually starts to move. While the acausality is
characteristically weak if $k$ is small, so that these boundary conditions are not
especially problematic for radio emissions with long wave trains, it is unclear whether
they really give a usefully defined theory that allows for arbitrary electromagnetic excitations.

Runaway excitations in a theory are most typically associated with energies that are not
bounded below, and this is also the case for the Lorentz-violating Chern-Simons theory.
With an arbitrary Chern-Simons term present, the purely electromagnetic part of the energy-momentum tensor
becomes~\cite{ref-carroll1}
\begin{equation}
\Theta^{\mu\nu}= -F^{\mu\alpha}F^{\nu}\,_{\alpha}+\frac{1}{4}g^{\mu\nu}
F^{\alpha\beta}F_{\alpha\beta} -\frac{1}{2}k_{AF}^{\nu}\epsilon^{\mu\alpha\beta\gamma}
F_{\beta\gamma}A_{\alpha}.
\end{equation}
The fact that this tensor is not symmetric is a consequence of the Lorentz violation.
The restriction that $k_{AF}$ be purely timelike simplifies the components of
$\Theta^{\mu\nu}$ somewhat; the energy density
($\Theta^{00}$), momentum density ($\Theta^{0j}$), and energy flux ($\Theta^{j0}$)
are
\begin{eqnarray}
{\cal E} & = & \frac{1}{2}\vec{E}^{2}+\frac{1}{2}\vec{B}^{2}-
k\vec{A}\cdot\vec{B} \\
\vec{{\cal P}} & = & \vec{E}\times\vec{B} \\
\vec{S} & = & \vec{E}\times\vec{B}-kA_{0}\vec{B}+k\vec{A}\times\vec{E},
\end{eqnarray}
respectively.
Except for the momentum density, these quantities are obviously not gauge invariant.
However, the total energy, found by integrating ${\cal E}$ over all space, is
gauge invariant. This may still not be obvious from the form of ${\cal E}$, but
because ${\cal E}$ (and, similarly, the $k_{AF}$ Lagrange density) changes by a total
derivative under a gauge transformation, the total energy does not depend on the gauge.

The instability of the theory is tied to another unusual property of the very
same $-k\vec{A}\cdot\vec{B}$ term in the energy density. This term is not
bounded below. This term may be made arbitrarily
negative by increasing the amplitude of the field $\vec{A}$ (and thus also $\vec{B}$).
For modes with small momenta $Q<|2k|$, the new term can be larger in magnitude than the usual
magnetic energy $\frac{1}{2}\vec{B}^{2}$. However, for shorter wavelength modes, the
additional derivative in the usual magnetic energy makes the $\frac{1}{2}\vec{B}^{2}$ term
dominant. Thus
the form of ${\cal E}$ not only reveals the existence of the instability but also clarifies
why it is restricted to the longest-wavelength modes of the theory.

With a purely timelike $k_{AF}$, the only change to Maxwell's equations is to the
Ampere-Maxwell Law,
\begin{equation}
\vec{\nabla}\times\vec{B}-\frac{\partial\vec{E}}{\partial t}=2k\vec{B}+\vec{J}
\end{equation}
(although the changes are a bit more complicated---involving $\vec{E}$ as well---if
$k_{AF}$ possesses a spacelike part). The magnetic field becomes a source for itself,
behaving like an effective current source $\vec{J}_{{\rm eff}}=2k\vec{B}$. For comparatively
simple source configurations, the Maxwell's equations may be solved---sometimes
exactly~\cite{ref-schober2}, but more typically as a power series in the small
Lorentz violation parameter $k$.

For the purpose of studying Cerenkov radiation, the natural source configuration to
consider is a pointlike charge moving with a uniform velocity $\vec{v}$. In the
Chern-Simons theory, since there are modes of the radiation field with arbitrarily
small phase velocities $\omega_{\pm}/Q$, it would be natural to expect Cerenkov
radiation. If a real moving charge lost energy and momentum through such radiation, it
would naturally slow down, which would further modify the radiation. However, any
change to the radiation that depends on the acceleration of the charge is not truly
Cerenkov radiation and will not be a part of our analysis.

Previously, the $\vec{E}$ and $\vec{B}$ fields of the lone moving charge have been explored
as dual power series in $k$ and $v$~\cite{ref-schober1}. The key simplification is that
the fields depend on position $\vec{r}$ and time $t$ only through the combination
$\vec{r}-\vec{v}t$. The fields are carried along uniformly with the moving charge, and
this allows the time derivatives in Maxwell's equations to be replaced with spatial
derivatives. Due to the symmetry properties of the field solutions, it
has been possible to demonstrate the surprising result that the Cerenkov power emitted by
the moving charge exactly vanishes, at all order in $k$. There is a
superficially reasonable explanation
for why the power can vanish: Modes of the field carrying negative energies can cancel the
energy carried by positive-energy, real-frequency modes. However, the cancelation is
fragile, and it does not need to occur if there are other modifications to the electromagnetic
sector, beyond the Chern-Simons term~\cite{ref-colladay3}. Moreover, while the inferred explanation
for the cancelation makes sense, it was arrived at without any study of the
behavior of the theory in Fourier space, on a mode-by-mode level.

The present work is aimed at providing that missing analysis. We will study the behavior
of the magnetic field and the outward energy flow using Fourier methods. Assigning physical
interpretations to various terms in this theory is always a bit tricky, because, as previously
noted, many quantities in the theory are not gauge invariant. For this analysis, we shall
choose to work in a single gauge---the Coulomb gauge $\vec{\nabla}\cdot\vec{A}=0$---because
it has been found to simplify the forms taken by the potentials $\vec{A}$ and $A_{0}$ a great deal.
(In particular, $A_{0}$ is completely independent of $k$ in this gauge.)

\section{Fourier Transforms of Field and Poynting Vector}
\label{sec-fourier}

The starting point for the present calculations will be the leading-order
$k$-dependent magnetic field
\begin{equation}
\label{eq-B11}
\vec{B}^{(1,1)}=\frac{kqv}{4\pi r}\left[\hat{v}+(\hat{v}\cdot\hat{r})\hat{r}\right],
\end{equation}
which was calculated in~\cite{ref-altschul36}; the superscripts indicate that this
is the magnetic field at first order in $k$ and in the speed $v$. We shall calculate
the Fourier transform of this magnetic field and also transforms of
other important functions in the theory (including some higher-order terms in
the magnetic field expansion).

Another function of particular importance is
the term $-kA_{0}\vec{B}$ that is part of the modified Poynting vector. In fact, at the
lowest orders in $v$, only this term can contribute to an outwardly directed energy
flux $\vec{S}\cdot\hat{r}$. The other two terms in $\vec{S}$ involve cross products with
the electric field $\vec{E}$; and since $\vec{E}$ points in the radial direction for
terms up to ${\cal O}(v)$, these terms cannot contribute to $\vec{S}\cdot\hat{r}$
below ${\cal O}(v^2)$.

Thanks to the simplicity of the Coulomb gauge, the scalar potential $A_{0}$ remains
\begin{equation}
A_{0}=\frac{q}{4\pi r\sqrt{1-v^{2}\left[1-\left(\hat{v}\cdot\hat{r}\right)^{2}\right]}}.
\end{equation}
To the order at which we are presently performing calculations, all that is needed
is $A_{0}^{(0,0)}=q/4\pi r$. From this it is clear that 
key to our analysis will be the Fourier transforms of functions of the general form
$r^{-n}[\hat{v}+(\hat{v}\cdot\hat{r})\hat{r}]$. In performing the necessary
transformations, the integrals involved may
require regularization at either large or small values of $r$.

For the Fourier transform, we shall select the usual coordinates, so that the $z$-axis
points along direction of the wave number variable $\vec{Q}$. Then the position
$\vec{r}$ and the velocity $\vec{v}$ will be expressed in spherical coordinates
with respect to this polar axis---$\vec{r}=(r,\theta,\phi)$ and
$\vec{v}=(v,\theta',\phi')$, respectively. Expressed in these coordinates, the
necessary Fourier transforms are
\begin{equation}
{\cal I}_{n}=\int d^{3}r \frac{e^{-\mu r}e^{i\vec{Q}\cdot\vec{r}}}{(r^{2}+\lambda^{2})^{n/2}}
\left[\hat{v}+\left(\hat{v}\cdot\hat{r}\right)\hat{r}\right].
\end{equation}
The quantities $\lambda$ and $\mu$ regularize the integral at small and large $r$,
respectively. Both will be taken to zero at the end of the calculation; however,
depending on the specific integral being considered, it may be possible to set one
or the other of them to zero earlier in the evaluation process.

Writing ${\cal I}_{n}$ as an iterated integral, we find
\begin{eqnarray}
{\cal I}_{n} & = & \int_{0}^{\infty}r^{2}\,dr\frac{e^{-\mu r}}{(r^{2}+\lambda^{2})^{n/2}}
\int_{0}^{\pi}\sin\theta\, d\theta\, e^{iQr\cos\theta}
\int_{0}^{2\pi}d\phi\,\left\{\hat{v}+\left[\sin\theta\sin\theta'\left(
\cos\phi\cos\phi' \right.\right.\right. \nonumber\\
& & \left.\left.\left. +\sin\phi\sin\phi'\right)+\cos\theta\cos\theta'\right]
\left(\sin\theta\cos\phi\,\hat{x}+\sin\theta\sin\phi\,\hat{y}+\cos\theta\,\hat{z}
\right)\right\}.
\end{eqnarray}
Only terms with even powers of $\cos\phi$ and $\sin\phi$ are nonzero under $\phi$ integration, so
\begin{eqnarray}
{\cal I}_{n} & = & \int_{0}^{\infty}dr\frac{r^{2}e^{-\mu r}}{(r^{2}+\lambda^{2})^{n/2}}
\int_{0}^{\pi}\sin\theta\, d\theta\, e^{iQr\cos\theta}
(\pi)\left[2\hat{v} \right. \\
& & \left. +\sin^{2}\theta\sin\theta'\left(\cos\phi'\,\hat{x}+
\sin\phi'\,\hat{y}\right)+2\cos^{2}\theta\cos\theta'\,\hat{z}\right]. \nonumber
\end{eqnarray}
Making the standard substitution $u=\cos\theta$,
\begin{equation}
\label{eq-u-sub}
{\cal I}_{n}=\pi\int_{0}^{\infty}dr\frac{r^{2}e^{-\mu r}}{(r^{2}+\lambda^{2})^{n/2}}
\int_{-1}^{1}du\,e^{iQru}\left[2\hat{v}+(1-u^{2})\sin\theta'\left(\cos\phi'\,\hat{x}+
\sin\phi'\,\hat{y}\right)+2u^{2}\cos\theta'\,\hat{z}\right].
\end{equation}
Using the elementary integrals
\begin{eqnarray}
\label{eq-u0}
\int_{-1}^{1}du\, e^{iQru} & = & \frac{2\sin Qr}{Qr} \\
\label{eq-u2}
\int_{-1}^{1}du\, u^{2}e^{iQru} & = & \frac{2\sin Qr}{Qr}+\frac{4\cos Qr}{(Qr)^{2}}
-\frac{4\sin Qr}{(Qr)^{3}},
\end{eqnarray}
the formula for ${\cal I}_{n}$ is reduced to a single radial integral
\begin{eqnarray}
\label{eq-In-final}
{\cal I}_{n} & = & 4\pi\int_{0}^{\infty}dr\frac{r^{2}e^{-\mu r}}{(r^{2}+\lambda^{2})^{n/2}}
\left\{\left(\hat{v}+\cos\theta'\,\hat{z}\right)\left(\frac{\sin Qr}{Qr}\right) \right. \\
& & \left. +\left[2\cos\theta'\,\hat{z}-\sin\theta'\left(\cos\phi'\,\hat{x}+
\sin\phi'\,\hat{y}\right)\right]\left[\frac{Qr\cos Qr-\sin Qr}{(Qr)^{3}}\right]\right\} \nonumber
\end{eqnarray}

With the general structure of ${\cal I}_{n}$ established, it remains to evaluate it for the 
cases of particular interest, which correspond principally to $n=1$ or 2. For $n=1$, which provides
the Fourier transform of $\vec{B}^{(1,1)}$, the regulation at small $r$ is unnecessary, and
$\lambda$ may be set to zero. Then first term in French brackets in (\ref{eq-In-final}) just
reproduces the usual integral that appears in the Fourier transform of a $r^{-1}$ potential.
The second term is only slightly more complicated, becoming
\begin{equation}
\frac{4\pi}{Q^{3}}\left(3\cos\theta'\,\hat{z}-\hat{v}\right)\int_{0}^{\infty}dr\frac{e^{-\mu r}
(Qr\cos Qr-\sin Qr)}{r^{2}}=
\frac{4\pi}{Q^{3}}\left(3\cos\theta'\,\hat{z}-\hat{v}\right)\left[-Q+\mu\tan^{-1}\left(\frac{Q}{\mu}
\right)\right].
\end{equation}
This makes the sum of the terms, as $\mu\rightarrow0$,
\begin{equation}
{\cal I}_{1}=\frac{8\pi}{Q^{2}}(\hat{v}-\cos\theta'\,\hat{z}).
\end{equation}
Recalling that $\hat{z}$ is the direction of $\vec{Q}$, we can express the Fourier transform of
$\vec{B}^{(1,1)}$ in a coordinate-independent fashion as
\begin{equation}
\label{eq-B11-final}
\widetilde{\vec{B}^{(1,1)}}=\frac{2kqv}{Q^{2}}\left[\hat{v}-\left(\hat{v}\cdot\hat{Q}\right)\hat{Q}\right].
\end{equation}

For the $n=2$ case, the first term in (\ref{eq-In-final})---the one with $(\sin Qr)/Qr$---requires
only $\mu$ to keep it finite. Taking $\lambda\rightarrow0$, we are left with
\begin{equation}
\frac{4\pi}{Q}\left(\hat{v}+\cos\theta'\,\hat{z}\right)\int_{0}^{\infty}dr\frac{e^{-\mu r}\sin Qr}{r}
=\frac{4\pi}{Q}\left(\hat{v}+\cos\theta'\,\hat{z}\right)\tan^{-1}\left(\frac{Q}{\mu}\right)
\rightarrow\frac{2\pi^{2}}{|Q|}\left(\hat{v}+\cos\theta'\,\hat{z}\right),
\end{equation}
where the last limit applies as $\mu\rightarrow0$.

The second term is finite without the $\mu$ regularization, but $\lambda$ is required to keep the
integration well defined. In this case, with $\mu=0$ the term is
\begin{equation}
\frac{4\pi}{Q^{2}}\left(3\cos\theta'\,\hat{z}-\hat{v}\right)\int_{0}^{\infty}dr\frac{\cos Qr-
\frac{\sin Qr}{Qr}}{r^{2}+\lambda^{2}}=\frac{2\pi^{2}}{\lambda^{2}Q^{3}}
\left(3\cos\theta'\,\hat{z}-\hat{v}\right)\left[|\lambda|Qe^{-|\lambda Q|}+\frac{|Q|}{Q}
\left(e^{-|\lambda Q|}-1\right)\right].
\end{equation}
Now taking $\lambda\rightarrow0$, this becomes
\begin{equation}
\frac{4\pi}{Q^{2}}\left(3\cos\theta'\,\hat{z}-\hat{v}\right)\int_{0}^{\infty}dr\frac{\cos Qr-
\frac{\sin Qr}{Qr}}{r^{2}+\lambda^{2}}\rightarrow
-\frac{\pi^{2}}{|Q|}\left(3\cos\theta'\,\hat{z}-\hat{v}\right).
\end{equation}
Taking the two terms together,
\begin{equation}
\label{eq-I2-final}
{\cal I}_{2}=\frac{\pi^{2}}{\left|\vec{Q}\right|}\left[3\hat{v}-\left(\hat{v}\cdot\hat{Q}\right)\hat{Q}\right],
\end{equation}
so that
\begin{equation}
\label{eq-kA0B}
-k\widetilde{A_{0}\vec{B}^{(1,1)}}=-\frac{k^{2}q^{2}v}{16Q}\left[3\hat{v}-\left(\hat{v}\cdot\hat{Q}\right)
\hat{Q}\right].
\end{equation}
This result can also be obtained by convolving the Fourier transforms of $A_{0}$ and $\vec{B}^{(1,1)}$;
see the appendix for details.

In this gauge (and to this order), the Fourier transform of the energy transport term
$-kA_{0}\vec{B}$ indicates that a mode with wave vector $\vec{Q}$ does appear to carry energy. The
term proportional to $\vec{v}/Q$ is derived from the first term in (\ref{eq-B11}), proportional
to $\vec{v}/r$. This represents an apparent flow of energy from
the direction the charge has come from, toward the direction the charge is going. This
corresponds to a similar energy flow that has been identified in coordinate space---which does not deposit
a net energy
anywhere. It is analogous to the constant Poynting vector $\vec{S}=\vec{E}\times\vec{B}$
that exists in the presence of uniform, crossed electric and magnetic fields; although in
this case it is also relevant that the Poynting vector itself is not a gauge invariant quantity.

The second, $\hat{Q}$-dependent term is a bit more subtle. However, it is still possible to see
that this term does not lead to any outflow of energy from the vicinity of the charge to spatial infinity.
This fact is actually implied by the symmetry properties of (\ref{eq-I2-final}); a Fourier space version
of the general symmetry argument from~\cite{ref-schober1} could be applied to demonstrate this.
However, we shall instead show the vanishing explicitly at this order.

The total radiating power from the moving charge is the integral of $\vec{S}\cdot\hat{r}$; this is
equivalent to a three-dimensional integral of $\vec{\nabla}\cdot\vec{S}$ over all space. In
Fourier space, this means the outflow of energy is proportional to an integral over all $\vec{Q}$ of
the dot product of the wave vector $\vec{Q}$ with the Fourier transform of $\vec{S}$. Since only
the $-kA_{0}\vec{B}$ term in $\vec{S}$ is capable of describing energy outflow in our chosen gauge,
the power radiated at this order must be proportional to
\begin{equation}
P\propto\int d^{3}Q\,\left[\frac{3\hat{v}-\left(\hat{v}\cdot\hat{Q}\right)\hat{Q}}{Q}\right]\cdot\vec{Q}
=\int d^{3}Q\,\left(2\hat{v}\cdot\hat{Q}\right)=0.
\end{equation}
So, although it is less obvious for the $\left(\hat{v}\cdot\hat{Q}\right)\hat{Q}$ than for the $\hat{v}$
term, each of these terms describes a distribution of energy among the Fourier modes that does not actually
represent radiation from the moving charge out to infinity.

In general, when $-k\widetilde{A_{0}\vec{B}^{(m,l)}}$ takes the form $X(Q,\theta')\hat{v}+Y(Q,\theta')\hat{Q}$,
where $\theta'$ is still the angle between $\vec{Q}$ and $\vec{v}$, the net energy outflow vanishes
if $X$ is an even function of $\cos\theta'=\hat{v}\cdot\hat{Q}$ and $Y$ is an odd function of $\cos\theta'$.
That way, the dot product of $\vec{Q}$ with either term is an odd function of $\cos\theta'$; when
integrated over all $\vec{Q}$, the result therefore vanishes. For the higher-order $\widetilde{\vec{B}^{(m,l)}}$
terms [which are ${\cal O}(k^{m}v^{l})$], this same symmetry argument always applies,
and this can be seen explicitly for the $\widetilde{\vec{B}^{(m,1)}}$
terms derived below in section~\ref{sec-cancel}. In Fourier space, the vanishing of the total
power is established based on whether individual terms are even or odd functions of $\hat{v}\cdot\hat{Q}$; and
this is very similar to how the cancelation argument proceeds in real space, where it is based on the parity
of the field components with respect to $\hat{v}\cdot\hat{r}$.

\section{Cancelations Between Low- and High-$Q$ Modes}
\label{sec-cancel}

The Fourier transforms calculated above (and their relationships to the energy flow)
are interesting on their own, although they are, in some sense, just translations of results that were
previously known in position space into Fourier space. However, with what we now understand of the
Fourier decomposition of the energy flow, it is possible to derive some further results that are not so
readily expressible in coordinate space.

In~\cite{ref-schober1}, inferences were drawn about the mechanism by which the excited radiation field
somehow manages to carry away zero net energy. These inferences were correct, but they were basically
qualitative. The gist was as follows: A phase space estimate of the energy carried away by the
$Q>|2k|$ modes of the field yields a positive result. This is not exactly wrong, but the
omission of the $Q<|2k|$ is a critical problem. Since (for the
troublesome helicity) those modes do not possess a dispersion relations with a real frequency, they are
not amenable to study using phase space methods. Yet they can still make key contributions to the
energy flow. Since the total power emitted by the charge is zero, the $Q<|2k|$ modes must
be carrying negative energy---in an amount which exactly cancels the energy carried by the shorter
$Q>|2k|$ modes. This may initially appear puzzling, since normally, the involvement of these
modes would be expected to lead to instabilities; their imaginary frequencies would give the field
an exponentially increasing time dependence. However, our framework adroitly manages to avoid
that difficulty; by studying field configurations in which the field profiles are in constant
uniform motion, we have forced the modes to behave as propagating modes.
Instead of growing exponentially, the unstable modes are associated with propagating
solutions carrying negative energies.

With our current understanding of the behavior of the fields and the Poynting vector in 
Fourier space, we are now better equipped to understand this cancelation. Yet there are
still subtleties to the analysis. In particular, the Fourier transforms we have found
so far are not sufficient to display the cancelation behavior. The transforms we have
calculated all depend on $k$ as simple powers. With this kind of $k$ dependence, it is
clearly not possible to have any cancelations between effects at small and large $Q$; whether
$|k|$ is greater than or less than $Q/2$ cannot affect the sign of a term with this form.
In order to find the cancelation between different $Q$ ranges, we must look at interference
between terms at different orders in $k$.

We will look specifically at all the magnetic field terms that are of the lowest (linear)
order in the speed $v$. The equations for such terms are
\begin{eqnarray}
\label{eq-Bm-curl}
\vec{\nabla}\times\vec{B}^{(m,1)} & = & 2k\vec{B}^{(m-1,1)} \\
\label{eq-Bm-div}
\vec{\nabla}\cdot\vec{B}^{(m,1)} & = & 0.
\end{eqnarray}
There are no contributions from $\partial\vec{E}/\partial t$, because a $k$-dependent $\vec{E}$ term
can itself only be generated by the time dependence of a $k$-dependent $\vec{B}$ term,
which makes the $\vec{E}$ term involved necessarily of higher order in $v$.
Iterating the curl equation (\ref{eq-Bm-curl}) and applying, as usual, the solenoidal field condition
(\ref{eq-Bm-div}) gives
\begin{equation}
\label{eq-Bm+2}
-\vec{\nabla}^{2}\vec{B}\,^{(m+2,1)}=(2k)^{2}\vec{B}^{(m,1)}.
\end{equation}
In Fourier space, this becomes simply $Q^{2}\widetilde{\vec{B}^{(m+2,1)}}=(2k)^{2}\widetilde{\vec{B}^{(m,1)}}$,
or, resumming all the terms with odd powers of $k$,
\begin{equation}
\label{eq-Bodd}
\widetilde{\vec{B}^{({\rm odd},1)}}=\frac{Q^{2}}{Q^{2}-(2k)^{2}}\widetilde{\vec{B}^{(1,1)}}
=\frac{2kqv}{Q^{2}-(2k)^{2}}\left[\hat{v}-\left(\hat{v}\cdot\hat{Q}\right)\hat{Q}\right].
\end{equation}
Now the difference of the signs for the Fourier modes with $Q$ above and below $|2k|$ is clearly manifest.
For each individual term $\widetilde{\vec{B}^{(m,1)}}$ with odd $m>0$, its contribution to the 
$-kA_{0}\vec{B}$ energy outflow vanishes for symmetry reasons, whether in coordinate space or
Fourier space. Viewed from this viewpoint, the nature of the cancelation between short- and
long-wavelength mode is obscure. However, when we combine terms of different orders in $k$,
we reveal a singularity and sign change at $Q=|2k|$, confirming the earlier inferences
about low- and high-$Q$ cancelations. The infinity in the Fourier transform is is not a problem in
this context, since a principal value integration through the pole at $Q=|2k|$ will always yield a finite result; and this
is just another facet of the cancelation between the short- and long-wavelength modes.

Note that the expression (\ref{eq-Bodd}) is necessarily an even function of
$Q$, because without knowing the sign of $k$, it is impossible to determine whether the
pole in $Q$ occurs at $2k$ or $-2k$. The denominator involving $Q^{2}$ automatically captures both possible
pole locations in a single expression.

\section{Conclusions}
\label{sec-concl}

The presence of the pole in (\ref{eq-Bodd}) is not, in retrospect, particularly surprising.
In the Chern-Simons theory, a static magnetic field in vacuum obeys the Helmholtz equation
$\left[\vec{\nabla}^{2}+(2k)^{2}\right]\vec{B}=0$. This leads to a screening of magnetostatic
fields, which provides another way of constraining $k$ experimentally---although the resulting
bounds are much weaker than those derived from cosmological birefringence measurements. With a
moving point charge, the magnetic fields are not truly static; however, by only considering
effects at ${\cal O}(v)$, we have effectively neglected the time dependence of $\vec{B}$.
The remaining field at lowest order in $v$ then satisfies equation (\ref{eq-Bm+2}), just as
does a time-independent vacuum field.

Nonetheless, (\ref{eq-Bodd}) is a significant result. As already noted, each individual magnetic
field $\vec{B}^{(m,1)}$, for odd $m$, makes a contribution to $\vec{S}$ that does not represent
any real energy outflow. This follows from symmetry arguments, but it is not very illuminating.
By summing up an infinite number of terms---each of which, on its own, gives a vanishing integrated
power---we have shown an explicit change in the sign of the outward Poynting vector at $Q=|2k|$.
This validates all the inferences that had previously been drawn about how and why the total
Cerenkov emission rate vanishes in the theory.

The pole and sign change at $Q=|2k|$ occur in the Fourier transform of a gauge-invariant
quantity, the magnetic field. However, the generalization of results such as (\ref{eq-kA0B}) depends
on the gauge. The gauge invariance of the total energy $\int d^{3}r\,{\cal E}$ is tied
to the fact that $\vec{\nabla}\cdot\vec{S}$ changes by
$-k\vec{\nabla}\cdot[\partial(\Lambda\vec{B})/\partial t]$ under a gauge transformation
with gauge function $\Lambda$. The fact that $\vec{S}$ is not gauge invariant on its own
makes assigning contributions to the Poynting vector precise interpretations impossible.
However, for a well-behaved gauge function $\Lambda(\vec{r}-\vec{v}t)$ that, like the
fields, moves along with the charge, the time derivative in
$-k\vec{\nabla}\cdot[\partial(\Lambda\vec{B})/\partial t]$ ensures that any changes to
the structure of $-k\widetilde{A_{0}\vec{B}^{(1,1)}}$ are of ${\cal O}(v^{2})$ or
higher.

We have only explored explicitly the Fourier transforms of the magnetic $\vec{B}^{(m,l)}$
terms with odd $m$ and $l=1$. However, the neglect of the even-$m$ terms $\vec{B}^{(m,1)}$
cannot affect the character of the energy outflow. At ${\cal O}(v)$, only the
$-kA_{0}\vec{B}$ term in $\vec{S}$ can represent a radial outflow,
but this requires $\vec{B}$ itself
to have a radial component. All the $\vec{B}^{(m,1)}$ terms with even $m$ are actually azimuthal,
pointing in the $\hat{\phi}$-direction; for example, 
\begin{equation}
\label{eq-B21}
\vec{B}^{(2,1)}=\frac{k^{2}qv}{2\pi}\sin\theta\,\hat{\phi}=\frac{k^{2}qv}{2\pi}
\left(\hat{v}\times\hat{r}\right),
\end{equation}
with Fourier transform (as calculated in the appendix)
\begin{equation}
\label{eq-B21-fourier}
\widetilde{\vec{B}^{(2,1)}}=\frac{4ik^{2}qv}{Q^{3}}\left(\hat{v}\times\hat{Q}\right).
\end{equation}
[In fact, all the $\vec{B}^{(m,l)}$ with even $m$ are azimuthal, regardless of $l$.]
These azimuthal terms cannot contribute to $\vec{S}\cdot\hat{r}$ at ${\cal O}(v)$.
Moreover, they still obey (\ref{eq-Bm+2}), so the
full expression for $\widetilde{\vec{B}^{({\rm even},1)}}$ can be determined just from
the Fourier transform of the usual leading-order field $\vec{B}^{(0,1)}$ of a moving
charge.

Analyses of the Fourier modes of the fields at ${\cal O}(v^{2})$ and higher might still
be somewhat interesting. However, at higher order in $v$, the calculations become much
more complicated, because of the additional involvement of the electric fields. There
are Lorentz-violating contributions to $\vec{E}$, which can in turn generate further $\vec{B}$
contributions
through the displacement current. Moreover, $\vec{E}$ field terms pointing in non-radial
directions may make direct contributions to the modified Poynting vector $\vec{S}$.

Performing any of these calculations at ${\cal O}(v^{2})$ and beyond
could be an interesting exercise,
but the Fourier transforms of higher-order terms appear extremely unlikely to provide any particular new
insights, because there does
not seem to be any reason to expect any qualitatively new features to their behavior.
There are additional novel contributions to $\vec{B}$ coming from both the new $k$-dependent source
term in the modified Ampere-Maxwell law and from the usual displacement current mechanism.
Any magnetic field term generated by $\partial\vec{E}/\partial t$ will then itself generate
an infinite series of terms involving higher powers of $k$. In Fourier space, each of these
separately generated sub-series can be summed as in (\ref{eq-Bodd}). This indicates that the
presence of the pole and sign change at $Q=|2k|$ are not limited to Fourier transforms at
${\cal O}(v)$.

Studies of Cerenkov radiation, whether they concern real radiation emitted by fast-moving
particles in matter or the theoretical possibility of Cerenkov emission in a
Lorentz-violating vacuum, are typically most easily undertaken in Fourier space. Whether
emission occurs is mostly determined by whether it is kinematically allowed for modes
with certain wave numbers. Mode-by-mode studies of the emission properties can be used
in many Lorentz-violating theories~\cite{ref-altschul12}, including the Chern-Simons
theory with a spacelike $k_{AF}$~\cite{ref-lehnert2}. However, the peculiar energetics of the timelike
$k_{AF}$ theory considered here make it impossible to apply the usual kind of mode
analysis directly. Instead, it has been necessary to study the shapes of the field profiles
in coordinate space. In this paper, we have taken the resulting field solutions
and transformed them explicitly into Fourier space.

As a result, we have addressed what may have been the last major puzzle associated with
the Cerenkov properties of the timelike Chern-Simons theory. While vanishing of the
radiated power is a consequence of the symmetries of the fields, in Fourier space
the vanishing can be recast as a cancelation. The Fourier transform of the magnetic
field changes sign at $Q=|2k|$, which confirms that the low-energy modes with imaginary
frequencies are carrying negative energies.

The knowledge of these Fourier transforms may have further interesting consequences
for how we understand Lorentz-violating and other unusual field theories. For example,
comparison with the Fourier decomposition of the excited modes in the spacelike
Chern-Simons theory may provide additional insights as to how the two theories are similar
as well as how they differ. Overall, this work provides further insight into how
the most exotic quantum field theories may behave.

\appendix

\section*{Appendix: Additional Fourier Integrals}

In this appendix, we present the calculation of the Fourier transform (\ref{eq-B21-fourier}) of
$\vec{B}^{(2,1)}$ and the alternative derivation of (\ref{eq-kA0B}) using a convolution.
For the first, we have (according to the same conventions describing the vectors
$\vec{r}$ and $\vec{v}$ in spherical coordinates that we used previously)
\begin{equation}
\label{eq-B21-calc}
\widetilde{\vec{B}^{(2,1)}}=\frac{k^{2}qv}{2\pi}\int_{0}^{\infty}r^{2}\,dr\,e^{-\mu r}
\int_{0}^{\pi}\sin\theta\,d\theta\,e^{iQr\cos\theta}\int_{0}^{2\pi}d\phi\,\left(\hat{v}\times\hat{r}\right).
\end{equation}
We have included the regularization factor $e^{-\mu r}$, to eliminate divergences at large $r$.
However, regularization at small $r$ is clearly unnecessary. There is no power law divergence
in (\ref{eq-B21}) in the vicinity of $r=0$, but there is still a singularity there, because of the
presence of the $\sin\theta$ factor in $\vec{B}^{(2,1)}$---$\theta$ being undefined at $r=0$.

Any term from (\ref{eq-B21-calc}) that is
linear in $\sin\phi$ or $\cos\phi$ will give zero after the $\phi$ integration.
In the cross product, this means contributions proportional to the $x$- and $y$-components
of $\hat{r}$ must vanish. The remaining $\phi$ integral is
\begin{equation}
\int_{0}^{2\pi}d\phi\,\left(\hat{v}\times\cos\theta\,\hat{z}\right)
=2\pi\cos\theta\left(\sin\theta'\sin\phi'\,\hat{x}-\sin\theta'\cos\phi'\,\hat{y}\right).
\end{equation}
This leaves the full Fourier transform (with the substitution $u=\cos\theta$) as
\begin{equation}
\widetilde{\vec{B}^{(2,1)}}=k^{2}qv\int_{0}^{\infty}dr\, r^{2}e^{-\mu r}\int_{-1}^{1}du\,
ue^{iQru}\left(\hat{v}\times\hat{Q}\right).
\end{equation}
Now the $u$ integration is elementary, as in (\ref{eq-u0}) and (\ref{eq-u2}), and we are left with
\begin{eqnarray}
\widetilde{\vec{B}^{(2,1)}} & = & \frac{2ik^{2}qv}{Q^{2}}\left(\hat{v}\times\hat{Q}\right)
\int_{0}^{\infty}dr\,\left(\sin Qr-Qr\cos Qr\right)e^{-\mu r} \\
& = & \frac{4ik^{2}qvQ}{(Q^{2}+\mu^{2})^{2}}\left(\hat{v}\times\hat{Q}\right) \\
\label{eq-B21-final}
& \rightarrow & \frac{4ik^{2}qv}{Q^{3}}\left(\hat{v}\times\hat{Q}\right).
\end{eqnarray}
where the last limit in (\ref{eq-B21-final}) obviously applies as $\mu\rightarrow0$,

With this result in hand, we can express the Fourier transform of the magnetic field
at ${\cal O}(v)$, to all orders in the Chern-Simons coefficient $k$,
\begin{equation}
\label{eq-Ball1}
\widetilde{\vec{B}^{({\rm all},1)}}=\frac{qv}{Q^{2}-(2k)^{2}}\left[2k\,\hat{v}-2k\left(\hat{v}\cdot\hat{Q}\right)
\hat{Q}+iQ\left(\hat{v}\times\hat{Q}\right)\right].
\end{equation}
When transformed back to position space, the pole at $Q=|2k|$ will lead to sign-changing
oscillations in field strength at large distances $r$. The keys to (\ref{eq-Ball1})
having the required form were that the
Fourier transforms (\ref{eq-B11-final}) and (\ref{eq-B21-fourier}) do not have zeroes at
$Q=|2k|$.

The convolution leading to (\ref{eq-kA0B}) is trickier. The required integral is
\begin{equation}
\widetilde{A_{0}\vec{B}^{(1,1)}}=\tilde{A}_{0}\left(\vec{Q}\right)*
\widetilde{\vec{B}^{(1,1)}}\left(\vec{Q}\right)
=\frac{1}{2\pi}\int d^{3}\ell\, \tilde{A}_{0}\left(\vec{Q}-\vec{\ell}\,\right)
\widetilde{\vec{B}^{(1,1)}}\left(\vec{\ell}\,\right).
\end{equation}
Since $\widetilde{\vec{B}^{(1,1)}}$ contains terms proportional to $Q^{-2}$
and $Q^{-2}\left(\hat{v}\cdot\hat{Q}\right)\hat{Q}$, the full convolution may be split into
two separate terms. Using the well-known Fourier transform of the nonrelativistic
$A_{0}$, which is also proportional to $Q^{-2}$, the first, slightly simpler term is
determined by
\begin{equation}
\label{eq-conv-int1}
\frac{1}{Q^{2}}*\frac{1}{Q^{2}}=\frac{1}{2\pi}\int_{0}^{\infty}\ell^{2}\,d\ell
\int_{0}^{\pi}\sin\vartheta\,d\vartheta\int_{0}^{2\pi}d\varphi
\frac{1}{Q^{2}+\ell^{2}-2Q\ell\cos\vartheta}\frac{1}{\ell^{2}}.
\end{equation}
The spherical coordinates of the integration variable $\vec{\ell}$ are
$(\ell,\vartheta,\varphi)$. The $\varphi$ integration is manifestly trivial. With
the substitution $\upsilon=\cos\vartheta$, the remaining two integrals are
\begin{equation}
\frac{1}{Q^{2}}*\frac{1}{Q^{2}}=\int_{0}^{\infty}d\ell
\int_{-1}^{1}d\upsilon\frac{1}{Q^{2}+\ell^{2}-2Q\ell\upsilon}.
\end{equation}
The key observation is that, since the integration over $\upsilon$ ranges over a region
that is symmetric about zero, the integrand of the outermost $\ell$ integration is an
even function of $\ell$. Changing the sign of $\ell$ changes the value of the integrand.
However, simultaneously changing the sign of $\upsilon$ returns the integrand to its
original value, and all values of $\upsilon$ between $-1$ and 1 are included in the
integration.
This means that the $\ell$ integration may be extended to run from
$-\infty$ to $\infty$ (and then halved). Doing this and then reversing the order of integrations gives
\begin{eqnarray}
\label{eq-int-reverse}
\frac{1}{Q^{2}}*\frac{1}{Q^{2}} & = & \frac{1}{2}\int_{-1}^{1}d\upsilon\int_{-\infty}^{\infty}
d\ell\frac{1}{(\ell-Q\upsilon)^{2}+(Q^{2}-Q^{2}\upsilon^{2})} \\
& = & \frac{1}{2}\int_{-1}^{1}d\upsilon
\left.\left[\frac{1}{Q\sqrt{1-\upsilon^{2}}}\tan^{-1}\left(\frac{l-Q\upsilon}
{Q\sqrt{1-\upsilon^{2}}}\right)\right]\right|_{-\infty}^{\infty} \\
& = & \frac{\pi}{2Q}\int_{-1}^{1}d\upsilon\frac{1}{\sqrt{1-\upsilon^{2}}} \\
\label{eq-conv1}
& = & \frac{\pi^{2}}{2Q}.
\end{eqnarray}
This accounts for part of the final term proportional to $\hat{v}$

For the convolution with $Q^{-2}\left(\hat{v}\cdot\hat{Q}\right)\hat{Q}$, what is required is
merely assembling the result from other calculational elements that have already been completed.
The $\varphi$ integration is more complicated than in (\ref{eq-conv-int1}), but it has already been
done in (\ref{eq-u-sub}),
\begin{equation}
\int_{0}^{\pi}d\varphi\,\left(\hat{v}\cdot\hat{\ell}\right)\hat{\ell}=\pi
\left[(1-\upsilon^{2})\left(\sin\theta'\cos\phi'\,\hat{x}+\sin\theta'\sin\phi'\,\hat{y}\right)
+2\upsilon^{2}\cos\theta'\,\hat{z}\right].
\end{equation}
The remaining integrations proceed as in (\ref{eq-int-reverse}). Thus we have,
again reversing the order of the iterated integrals,
\begin{eqnarray}
\frac{1}{Q^{2}}*\frac{\left(\hat{v}\cdot\hat{Q}\right)\hat{Q}}{Q^{2}} & = &
\frac{1}{4}\int_{-1}^{1}d\upsilon\,\left[(1-\upsilon^{2})
\left(\sin\theta'\cos\phi'\,\hat{x}+\sin\theta'\sin\phi'\,\hat{y}-2\cos\theta'\,\hat{z}\right)\right.
\nonumber\\
& & +\left.2\cos\theta'\,\hat{z}\right]\int_{-\infty}^{\infty}
d\ell\frac{1}{(\ell-Q\upsilon)^{2}+(Q^{2}-Q^{2}\upsilon^{2})} \\
& =  & \frac{\pi}{4Q}\int_{-1}^{1}d\upsilon\,\left\{\sqrt{1-\upsilon^{2}}\left[
\hat{v}-3\left(\hat{v}\cdot\hat{Q}\right)\hat{Q}\right]+\frac{2\left(\hat{v}\cdot\hat{Q}\right)\hat{Q}}
{\sqrt{1-\upsilon^{2}}}\right\} \\
\label{eq-conv2}
& = & \frac{\pi^{2}}{8Q}\left[\hat{v}+\left(\hat{v}\cdot\hat{Q}\right)\hat{Q}\right].
\end{eqnarray}
Inserting the proper multiplicative factors
and taking a difference of (\ref{eq-conv1}) and (\ref{eq-conv2}),
we recover the result (\ref{eq-kA0B}).

\end{document}